\documentclass{article}

\usepackage{a4wide}

\usepackage{hyperref}
\usepackage{amsmath}
\usepackage{amssymb}
\usepackage{amsthm}
\usepackage{graphicx}
\usepackage{caption}
\usepackage{subcaption}

\newtheorem{definition}{Definition}

\begin{document}

\title{Sensitivity Analysis in a Dengue Epidemiological Model\thanks{Submitted 14-June-2013; 
accepted, after a minor revision, 30-June-2013; Conference Papers in Mathematics, 
Volume 2013, Article ID 721406, \url{http://dx.doi.org/10.1155/2013/721406}}}

\author{Helena Sofia Rodrigues$^{1,2,3}$\\ sofiarodrigues@esce.ipvc.pt
\and M. Teresa T. Monteiro$^{2}$\\ tm@dps.uminho.pt
\and Delfim F. M. Torres$^{3}$\\ delfim@ua.pt}

\date{$^1$School of Business, Polytechnic Institute of Viana do Castelo, Valen\c{c}a, Portugal\\[0.2cm]
$^2$R\&D Centre Algoritmi, Department of Production and Systems,\\ University of Minho, Braga, Portugal\\[0.2cm]
$^3$R\&D Centre CIDMA, Department of Mathematics,\\ University of Aveiro, Aveiro, Portugal}

\maketitle

% ---------------------------------------

\begin{abstract}
Epidemiological models may give some basic guidelines for public health practitioners,
allowing to analyze issues that can influence the strategies to prevent and fight a disease.
To be used in decision-making, however, a mathematical model must be carefully
parameterized and validated with epidemiological and entomological data.
Here a SIR (S for susceptible, I for infectious, R for recovered individuals)
and ASI (A for the aquatic phase of the mosquito, S for susceptible and I for infectious mosquitoes)
epidemiological model describing a dengue disease is presented,
as well as the associated basic reproduction number. A sensitivity analysis
of the epidemiological model is performed in order to determine the relative
importance of the model parameters to the disease transmission.
\end{abstract}

\begin{quotation}
\noindent \textbf{Keywords}: sensitivity analysis; basic reproduction number; epidemiological model; dengue.
\end{quotation}

% ---------------------------------------

\section{Introduction}

Dengue is a major public health problem in tropical and sub-tropical countries. It is
a vector-borne disease transmitted by \emph{Aedes aegypti} and \emph{Aedes albopictus}
mosquitoes. Four different serotypes can cause dengue fever. A human infected by
one serotype, when recovers, gains total immunity to that serotype,
and only partial and transient immunity with respect to the other three.

Dengue can vary from mild to severe. The more severe forms of dengue include
shock syndrome and dengue hemorrhagic fever (DHF). Patients who develop these
more serious forms of dengue fever usually need to be hospitalized.

The full life cycle of dengue fever virus involves the role of the mosquito as a transmitter
(or vector) and humans as the main victim and source of infection. Preventing or
reducing dengue virus transmission depends entirely in the control of
mosquito vectors or interruption of human-vector contact \cite{Who2009}.

In Section~\ref{sec:2} an epidemiological model for dengue disease is presented.
It consists of six mutually-exclusive compartments, expressing the
interaction between human and mosquito, and designed for examining
the process of the disease spread into a population.

Similarly to humans, mosquitoes differ among themselves in terms of their life history
traits. Besides individual variations, the environment (temperature and humidity)
also has strong effect on the life history \cite{Chen2012}. Another source
of uncertainties, regarding appropriate parameter values, is the scarcity of the data
available for the  mosquito population, and the diversity among the international data.

Our model includes a set of parameters related to human and  mosquito
populations and their interaction. Often the unknown parameters
involved in the models are assumed to be constant over time. However, in a more realistic
perspective of any phenomenon, some of them are not constant and implicitly
depend on several factors. Many of such factors usually do not appear explicitly
in the mathematical models because of the need of balance between modeling
and numerical tractability and the lack of a precise knowledge of them \cite{Luz2003}.

Sensitivity analysis allows to investigate how uncertainty
in the input variables affects the model outputs and which input variables tend
to drive variation in the outputs.
Sensitivity of the basic reproduction number for a tuberculosis model
can be found in \cite{MyID:271}. Here one of the goals is to determine
which parameters are worth pursuing in the field in order to develop a dengue
transmission model. For our specific model, a sensitivity analysis is performed
in Section~\ref{subsection_sensitivity} to determine the relative importance of the model parameters
to disease transmission, taking into account the basic reproduction number (Section~\ref{sec:3}).

Section~\ref{sec:4} reports some numerical experiments: a set of simulations
is presented to illustrate the effect of the parameters on the number
of infected individuals. Finally, some conclusions are given in Section~\ref{sec:5}.

% ---------------------------------------

\section{Dengue Model}
\label{sec:2}

Taking into account the model presented in
\cite{Dumont2010,Dumont2008} and the considerations
of \cite{Sofia2009,Sofia2010c}, a new model more adapted
to the dengue reality is proposed. The
notation used in the mathematical model includes three
epidemiological states for humans:
\begin{quote}
\begin{tabular}{l}
$S_h(t)$ --- susceptible (individuals who can contract the disease),\\
$I_h(t)$ --- infected (individuals capable of transmitting the disease),\\
$R_h(t)$ --- resistant (individuals who have acquired immunity).
\end{tabular}
\end{quote}
It is assumed that the total human population, $N_h$,
is constant: $N_h=S_h(t)+I_h(t)+R_h(t)$ at any time $t$.
The population is homogeneous, which means that every individual
of a compartment is homogeneously mixed with the other individuals.
Immigration and emigration are not considered.

Three other state variables, related
to the female mosquitoes, are considered:
\begin{quote}
\begin{tabular}{l}
$A_m(t)$ --- aquatic phase (that includes the egg, larva and pupa stages),\\
$S_m(t)$ --- susceptible (mosquitoes that are able to contract the disease),\\
$I_m(t)$ --- infected (mosquitoes capable of transmitting the disease).\\
\end{tabular}
\end{quote}
Note that male mosquitoes are not taken into account, because they are not capable
of transmitting the disease, and that there is no resistant phase,
due to the short lifespan of mosquitoes.

It is assumed homogeneity
between host and vector populations, which means that
each vector has an equal probability to bite any host.
Humans and mosquitoes are assumed to be born susceptible.
The dengue epidemic is modeled by the following nonlinear
system of time-varying ODEs (ordinary differential equations):
\begin{equation}
\label{cap6_ode1}
\begin{cases}
\displaystyle\frac{dS_h}{dt} = \mu_h N_h - \left(B\beta_{mh}\frac{I_m}{N_h}+\mu_h\right)S_h,\\[0.25cm]
\displaystyle\frac{dI_h}{dt} = B\beta_{mh}\frac{I_m}{N_h}S_h -(\eta_h+\mu_h) I_h,\\[0.25cm]
\displaystyle\frac{dR_h}{dt} = \eta_h I_h - \mu_h R_h,
\end{cases}
\end{equation}
and
\begin{equation}
\label{cap6_ode2}
\begin{cases}
\displaystyle\frac{dA_m}{dt} = \varphi \left(1-\frac{A_m}{k N_h}\right)(S_m+I_m)
-\left(\eta_A+\mu_A \right) A_m,\\[0.25cm]
\displaystyle\frac{dS_m}{dt} = \eta_A A_m
-\left(B \beta_{hm}\frac{I_h}{N_h}+\mu_m \right) S_m,\\[0.25cm]
\displaystyle\frac{dI_m}{dt} = B \beta_{hm}\frac{I_h}{N_h}S_m
-\mu_m I_m,
\end{cases}
\end{equation}
with initial conditions
\begin{equation}
\label{eq:IC}
\begin{tabular}{llll}
$S_h(0)=S_{h0}$, &  $I_h(0)=I_{h0}$, &
$R_h(0)=R_{h0}$, \\
$A_m(0)=A_{m0}$, & $S_{m}(0)=S_{m0}$, & $I_m(0)=I_{m0}$.
\end{tabular}
\end{equation}
The meaning of the parameters of the model,
together with the baseline values used in Section~\ref{subsection_sensitivity},
are given in Table~\ref{parameters}.
\begin{table}
\begin{center}
\begin{tabular}{lll}
\hline
\textbf{Parameter} & \textbf{Description} & \textbf{Value}\\
\hline
$N_h$ & total human population & 480000\\
$B$ & average daily biting (per day) & 0.8\\
$\beta_{mh}$ & transmission probability from $I_m$ (per bite)  & 0.375\\
$\beta_{hm}$ & transmission probability from $I_h$ (per bite) & 0.375\\
$\mu_{h}$ & average lifespan of humans (in days) & $1/(71\times365)$\\
$\eta_{h}$ & mean viremic period (in days)& $1/3$\\
$\mu_{m}$ & average lifespan of adult mosquitoes (in days) & $1/10$ \\
$\varphi$ & number of eggs at each deposit per capita (per day) & 6\\
$\mu_{A}$ & natural mortality of larvae (per day) & $1/4$\\
$\eta_{A}$ & maturation rate from larvae to adult (per day)& 0.08 \\
$m$ & female mosquitoes per human & 3\\
$k$ & number of larvae per human & 3\\
\hline
\end{tabular}
\caption{Parameters used in the dengue mathematical
model \eqref{cap6_ode1}--\eqref{cap6_ode2}.}
\label{parameters}
\end{center}
\end{table}

% ---------------------------------------

\section{Basic Reproduction Number}
\label{sec:3}

Due to biological reasons, only nonnegative solutions of the
initial value problem \eqref{cap6_ode1}--\eqref{eq:IC}
are acceptable. More precisely, it is necessary to study the solution properties
of the system \eqref{cap6_ode1}--\eqref{cap6_ode2} subject to given
initial conditions \eqref{eq:IC} in the closed set
\begin{equation*}
\Omega=\left\{(S_h,I_h,R_h,A_m,S_m,I_m)\in \mathbb{R}_{+}^{6}:
S_h+I_h+R_h\leq N_h,\,  A_m\leq k N_h, \, S_m+I_m \leq mN_h\right\}.
\end{equation*}
It can be verified that $\Omega$ is a positively invariant set
with respect to \eqref{cap6_ode1}--\eqref{cap6_ode2}. The proof
of this statement is similar to the one in \cite{Sofia2012}.

\begin{definition}
A sextuple $E = \left(S_h,I_h,R_h,A_m,S_m,I_m\right)$
is said to be an \emph{equilibrium point} for system
\eqref{cap6_ode1}--\eqref{cap6_ode2}
if it satisfies the following relations:
\begin{equation*}
\label{equilibrio}
\begin{cases}
\mu_h N_h - \left(B\beta_{mh}\frac{I_m}{N_h}+\mu_h\right)S_h=0,\\
B\beta_{mh}\frac{I_m}{N_h}S_h -(\eta_h+\mu_h) I_h=0,\\
\eta_h I_h - \mu_h R_h=0,\\
\varphi \left(1-\frac{A_m}{\alpha k N_h}\right)(S_m+I_m)-(\eta_A+\mu_A+c_A) A_m=0,\\
\eta_A A_m - \left(B \beta_{hm}\frac{I_h}{N_h}+\mu_m + c_m\right) S_m=0,\\
B \beta_{hm}\frac{I_h}{N_h}S_m -(\mu_m + c_m) I_m=0.
\end{cases}
\end{equation*}
An equilibrium point $E$ is \emph{biologically meaningful}
if and only if $E \in \Omega$. The biologically meaningful equilibrium points
are said to be disease free or endemic, depending on $I_h$ and $I_m$:
if there is no disease for both populations of humans and mosquitoes, that is,
if $I_h=I_m=0$, then the equilibrium point is said to be a
\emph{Disease Free Equilibrium} (DFE); otherwise, if $I_h \ne 0$ or $I_m \ne 0$,
in other words, if $I_h > 0$ or $I_m > 0$, then the equilibrium point is called \emph{endemic}.
\end{definition}

It is easily seen that system \eqref{cap6_ode1}--\eqref{cap6_ode2}
admits at most two DFE points. Let
$$
\mathcal{M} =-\left(\eta_A \mu_m+\mu_A \mu_m-\varphi \eta_A\right).
$$
If $\mathcal{M} \le 0$, then there is only
one biologically meaningful equilibrium point $E_{1}^{*}$:
$$
E_{1}^{*}=\left(N_h,0,0,0,0,0\right).
$$
If $\mathcal{M} > 0$, then there are two biologically meaningful disease free equilibrium points:
$E_{1}^{*}$ and
$$
E_{2}^{*}=\left(N_h,0,0,\frac{k N_h \mathcal{M}}{\eta_A\varphi},
\frac{k N_h
\mathcal{M}}{\mu_m \varphi},0\right).
$$
By algebraic manipulation, $\mathcal{M}>0$ is equivalent to condition
$$
\frac{(\eta_A+\mu_A)\mu_m}{\varphi\eta_A} < 1,
$$
which is related to the basic offspring number for mosquitoes:
if $\mathcal{M} \leq 0$, then the mosquito population will die out;
if $\mathcal{M} > 0$, then the mosquito population is sustainable,
and the equilibrium $E_{2}^{*}$ is more realistic
from a biological standpoint.

An important measure of transmissibility of the disease
is the epidemiological concept of
\emph{basic reproduction number} \cite{Heffernan2005}.
It provides an invasion criterion for the initial spread
of the virus in a susceptible population.

\begin{definition}
The basic reproduction number,
denoted by $\mathcal{R}_0$, is defined as the
average number of secondary infections that occurs when one
infective is introduced into a completely susceptible population.
\end{definition}

Using the next generation matrix of an ODE \cite{Driessche2002},
one concludes that the basic reproduction number $\mathcal{R}_0$
associated to the differential system
\eqref{cap6_ode1}--\eqref{cap6_ode2} is given,
in the case $\mathcal{M}>0$, by
\begin{equation}
\label{eqR0}
\mathcal{R}_{0}^{2} = \frac{k B^2 \beta_{hm} \beta_{mh}
\mathcal{M}}{\varphi (\eta_h + \mu_h)\mu_m^2}.
\end{equation}
If $\mathcal{R}_0 < 1$, then the disease
cannot invade the population and the infection will die out over
a period of time. The amount of time this will take generally depends on how
small $\mathcal{R}_0$ is. If $\mathcal{R}_0 > 1$, then an invasion is possible
and infection can spread through the population. Generally,
the larger the value of $\mathcal{R}_0$, the more severe, and possibly widespread,
the epidemic will be \cite{Sofia2013b}.

In determining how best to reduce human mortality and morbidity due to dengue,
it is necessary to know the relative importance of the different factors
responsible for its transmission. In the next section the sensitivity indices
of $\mathcal{R}_0$,  related to the parameters in the model, are calculated.

% ---------------------------------------

\section{Sensitivity Analysis}
\label{subsection_sensitivity}

Sensitivity analysis tell us how important each parameter is to
disease transmission. Such information is crucial not only for experimental design, but also
to data assimilation and reduction of complex nonlinear models \cite{Powell2005}.
Sensitivity analysis is commonly used to determine
the robustness of model predictions to parameter values, since there are usually errors in
data collection and presumed parameter values. It is used to discover parameters
that have a high impact on $\mathcal{R}_0$ and should be targeted by intervention strategies.

Sensitivity indices allow us to measure the relative change in a variable when a
parameter changes. The normalized forward sensitivity index of a variable with respect
to a parameter is the ratio of the relative change in the variable to the relative change in the parameter.
When the variable is a differentiable function of the parameter, the sensitivity index may
be alternatively defined using partial derivatives.

\begin{definition}[cf. \cite{Chitnis2008}]
The normalized forward sensitivity index of $\mathcal{R}_0$,
that depends differentiably on a parameter $p$, is defined by
$$
\Upsilon_p^{\mathcal{R}_0}
=\frac{\partial \mathcal{R}_0}{\partial p}\times \frac{p}{\mathcal{R}_0}.
$$
\end{definition}

Given the explicit formula \eqref{eqR0} for $\mathcal{R}_0$, one can easily
derive an analytical expression for the sensitivity of  $\mathcal{R}_0$ with
respect to each parameter that comprise it.
The obtained values are described in Table~\ref{sensitivity},
which presents the sensitivity indices for the baseline parameter
values in the last column of Table~\ref{parameters}.
Note that the sensitivity index may be a complex expression, depending on different parameters
of the system, but can also be a constant value, not depending on any of the parameter values.
For example, $\Upsilon_{\beta_{mh}}^{\mathcal{R}_0} \equiv +0.5$, meaning that
increasing (or decreasing) $\beta_{mh}$ by $10\%$ increases (or decreases) always
$\mathcal{R}_0$ by $5\%$.
\begin{table}
\begin{center}
\begin{tabular}{ll}
\hline
\textbf{Parameter} & \textbf{Sensitivity}\\
& \textbf{index}\\
\hline
$B$ & +1\\
$\beta_{mh}$ & +0.5 \\
$\beta_{hm}$ & +0.5\\
$\mu_{h}$ &  -0.0000578748\\
$\eta_{h}$ & -0.499942\\
$\mu_{m}$ & -1.03691 \\
$\varphi$ &  +0.0369128\\
$\mu_{A}$ &  -0.0279642\\
$\eta_{A}$ &  +0.527964\\
$k$ & +0.5 \\
\hline
\end{tabular}
\caption{Sensitivity indices of $\mathcal{R}_{0}$ evaluated at the
baseline parameter values given in Table~\ref{parameters}.}
\label{sensitivity}
\end{center}
\end{table}

A highly sensitive parameter should be carefully estimated, because a small
variation in that parameter will lead to large quantitative changes.
An insensitive parameter, on the other hand, does not require as much effort to estimate,
since a small variation in that parameter will not produce large changes to the quantity
of interest \cite{Mikucki2012}.

% ---------------------------------------

\section{Numerical Analysis}
\label{sec:4}

The simulations were carried out using the following values
for the initial conditions \eqref{eq:IC}:
\begin{equation}
\label{eq:ic}
\begin{gathered}
S_{h0}=N_h-10, \quad I_{h0}=10, \quad R_{h0}=0,\\
A_{m0}=k N_h, \quad S_{m0}=m N_h, \quad I_{m0}=0.
\end{gathered}
\end{equation}
The final time was $t_f=100$ days.
Computations were run in \textsf{Matlab} with the \texttt{ode45} routine.
This function implements a Runge--Kutta method with a variable time step for
efficient computation.

\begin{figure}
\centering
\begin{subfigure}[b]{0.48\textwidth}
\centering
\includegraphics[scale=0.53]{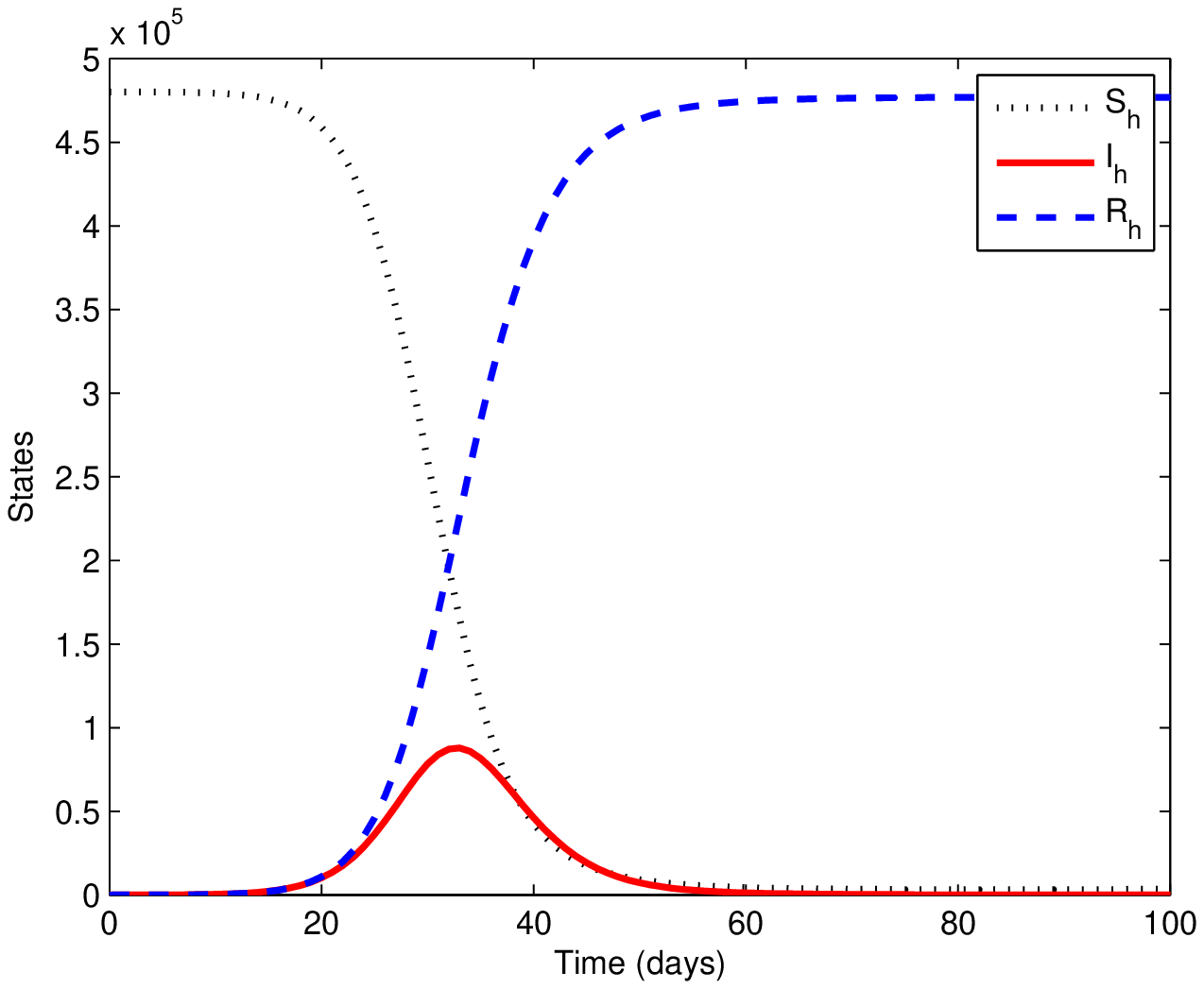}
\caption{Human population.}
\label{human}
\end{subfigure}
\begin{subfigure}[b]{0.48\textwidth}
\centering
\includegraphics[scale=0.53]{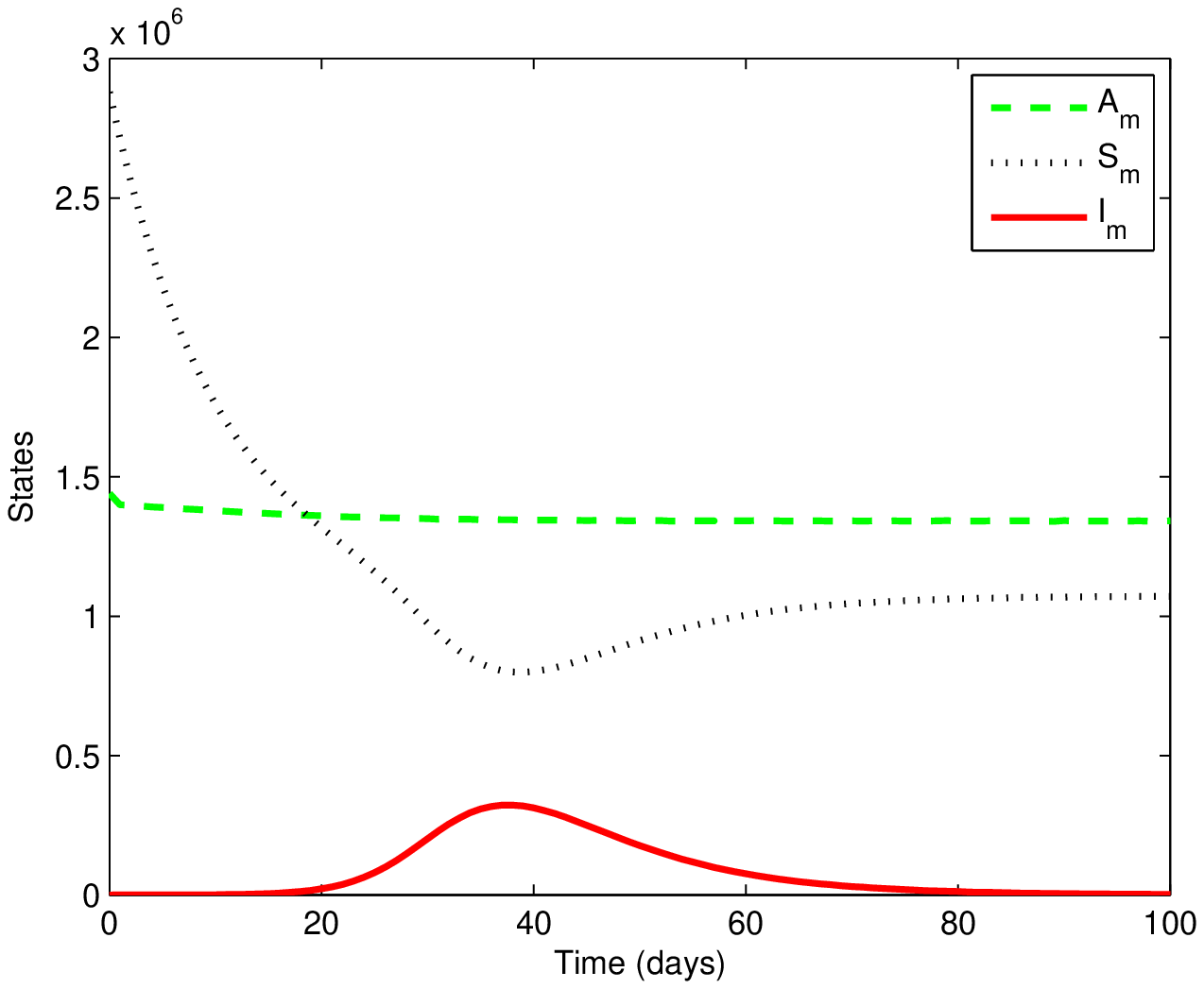}
\caption{Mosquito population.}
\label{mosquito}
\end{subfigure}
\caption{State variables of the ODE system \eqref{cap6_ode1}--\eqref{cap6_ode2}
with initial conditions \eqref{eq:ic} and parameters as in Table~\ref{parameters}.}
\label{population}
\end{figure}

Figures~\ref{human} and \ref{mosquito} show the solutions
to \eqref{cap6_ode1}--\eqref{eq:IC} with the baseline parameter values given
in Table~\ref{parameters}, for human and mosquitoes, respectively.

\begin{figure}
\centering
\begin{subfigure}[b]{0.45\textwidth}
\centering
\includegraphics[scale=0.44]{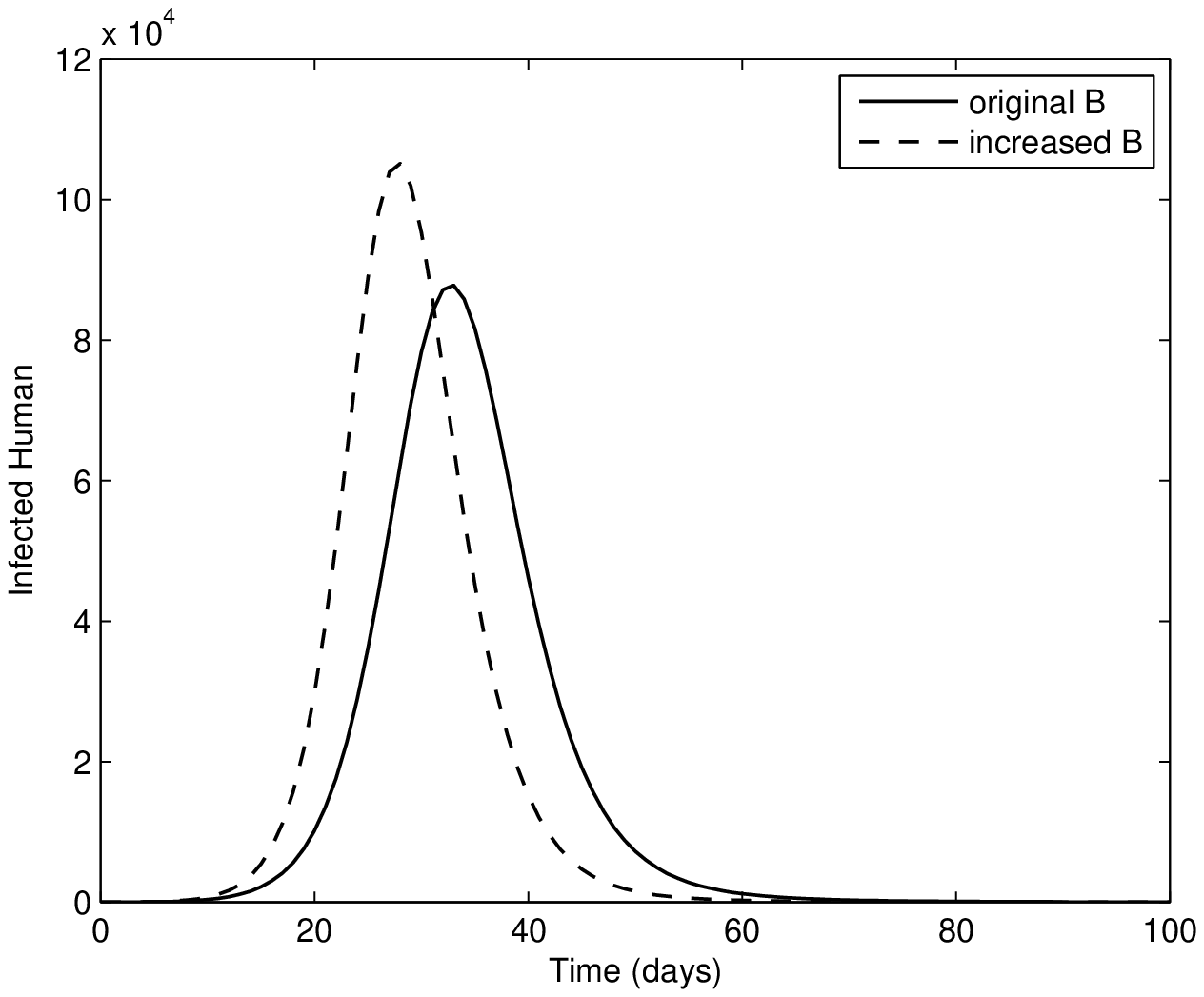}
\caption{Effect on $I_h$ of the variation of $B$.}
\label{B}
\end{subfigure}
\begin{subfigure}[b]{0.45\textwidth}
\centering
\includegraphics[scale=0.44]{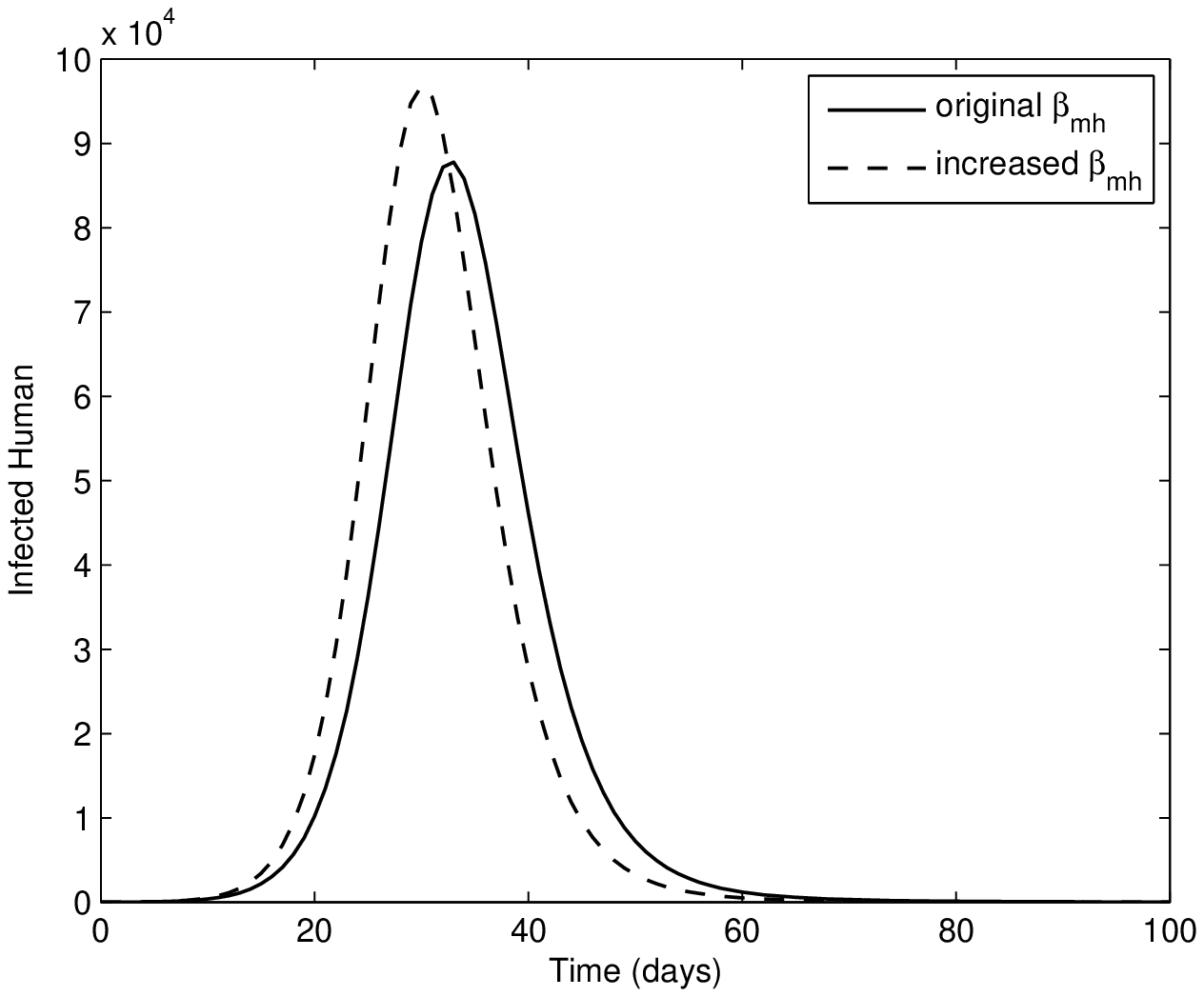}
\caption{Effect on $I_h$ of the variation of $\beta_{mh}$.}
\label{beta_mh}
\end{subfigure}
\begin{subfigure}[b]{0.45\textwidth}
\centering
\includegraphics[scale=0.44]{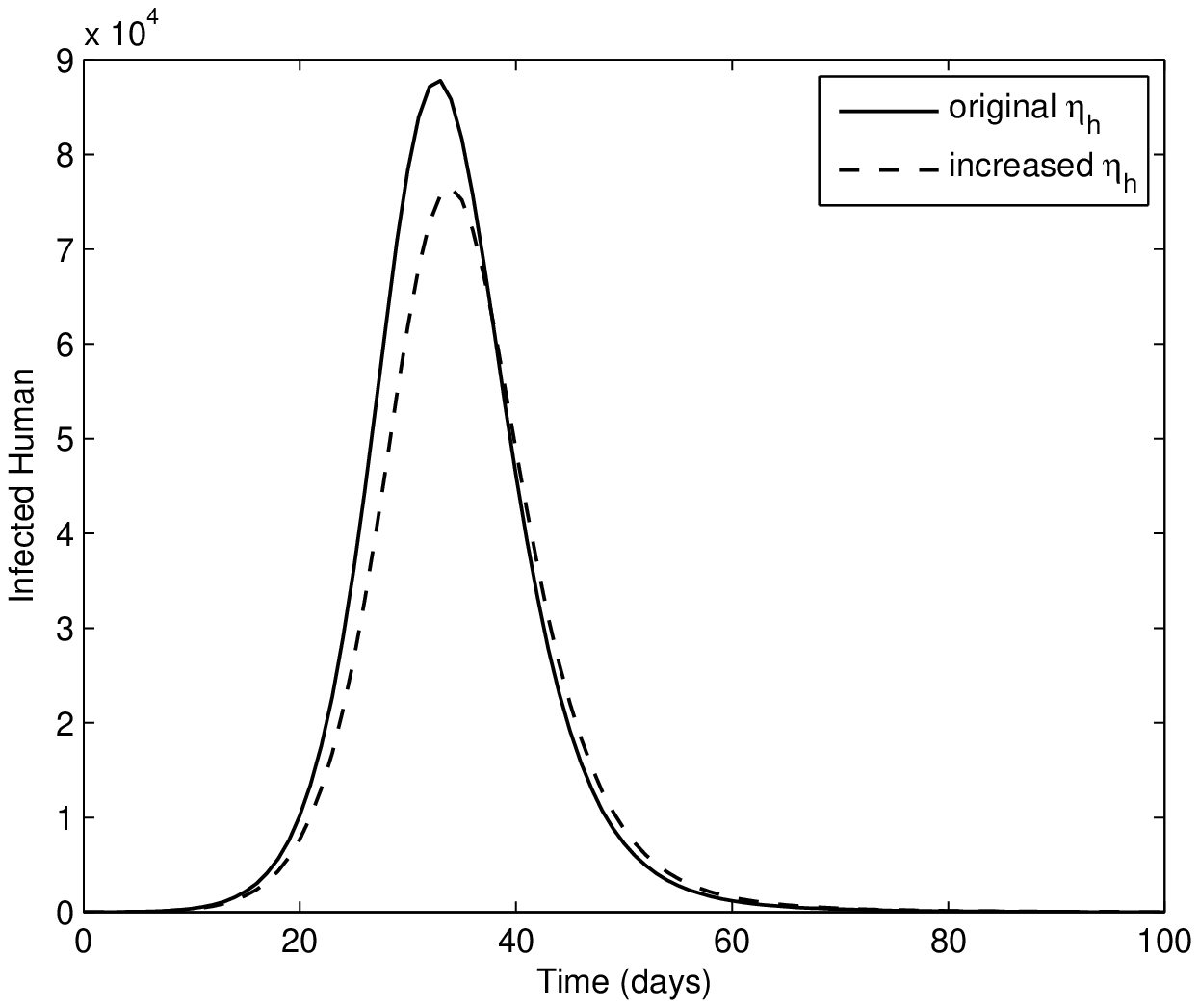}
\caption{Effect on $I_h$ of the variation of $\eta_h$.}
\label{eta_h_increased}
\end{subfigure}
\begin{subfigure}[b]{0.45\textwidth}
\centering
\includegraphics[scale=0.44]{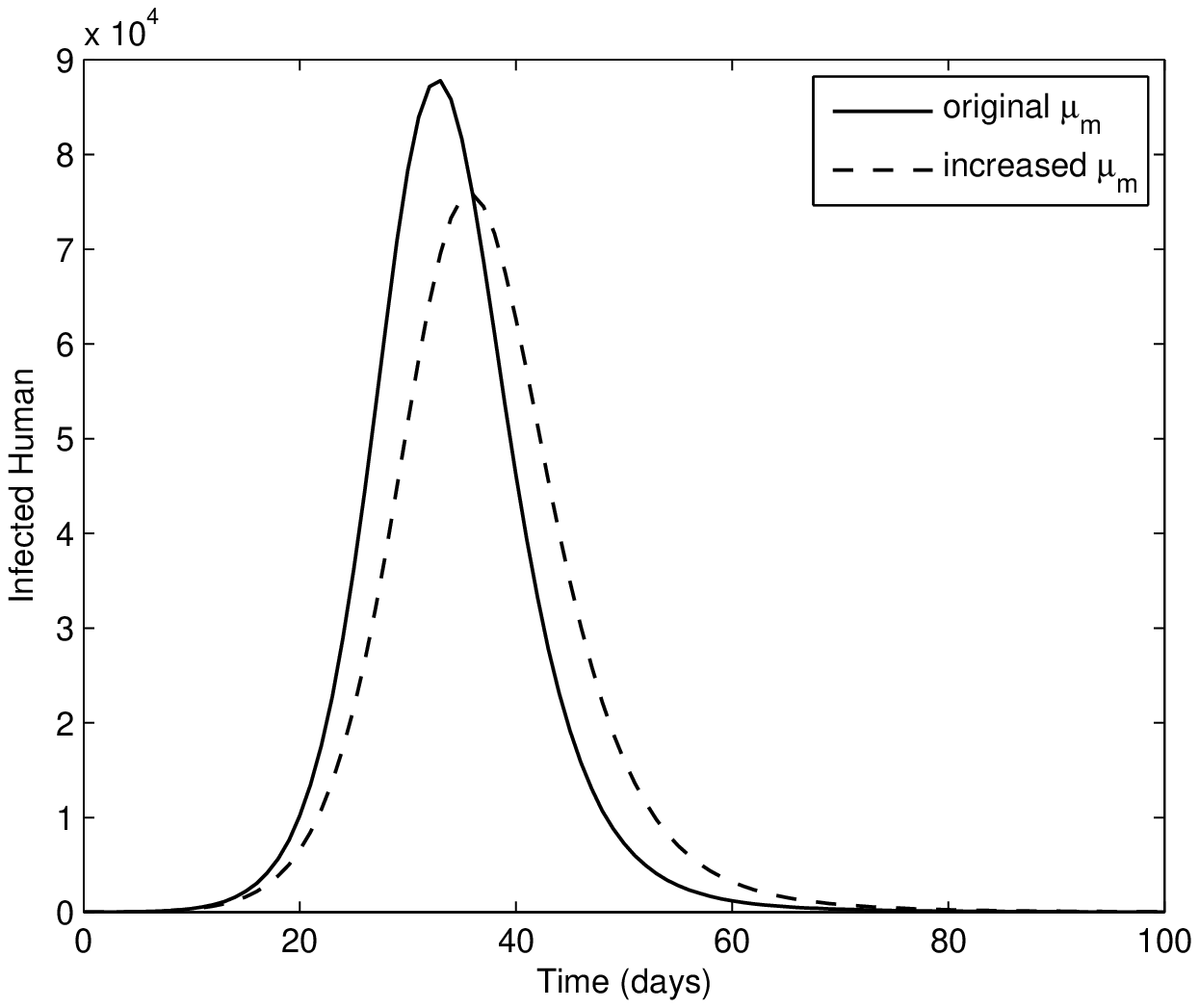}
\caption{Effect on $I_h$ of the variation of $\mu_m$.}
\label{mu_m}
\end{subfigure}
\begin{subfigure}[b]{0.45\textwidth}
\centering
\includegraphics[scale=0.44]{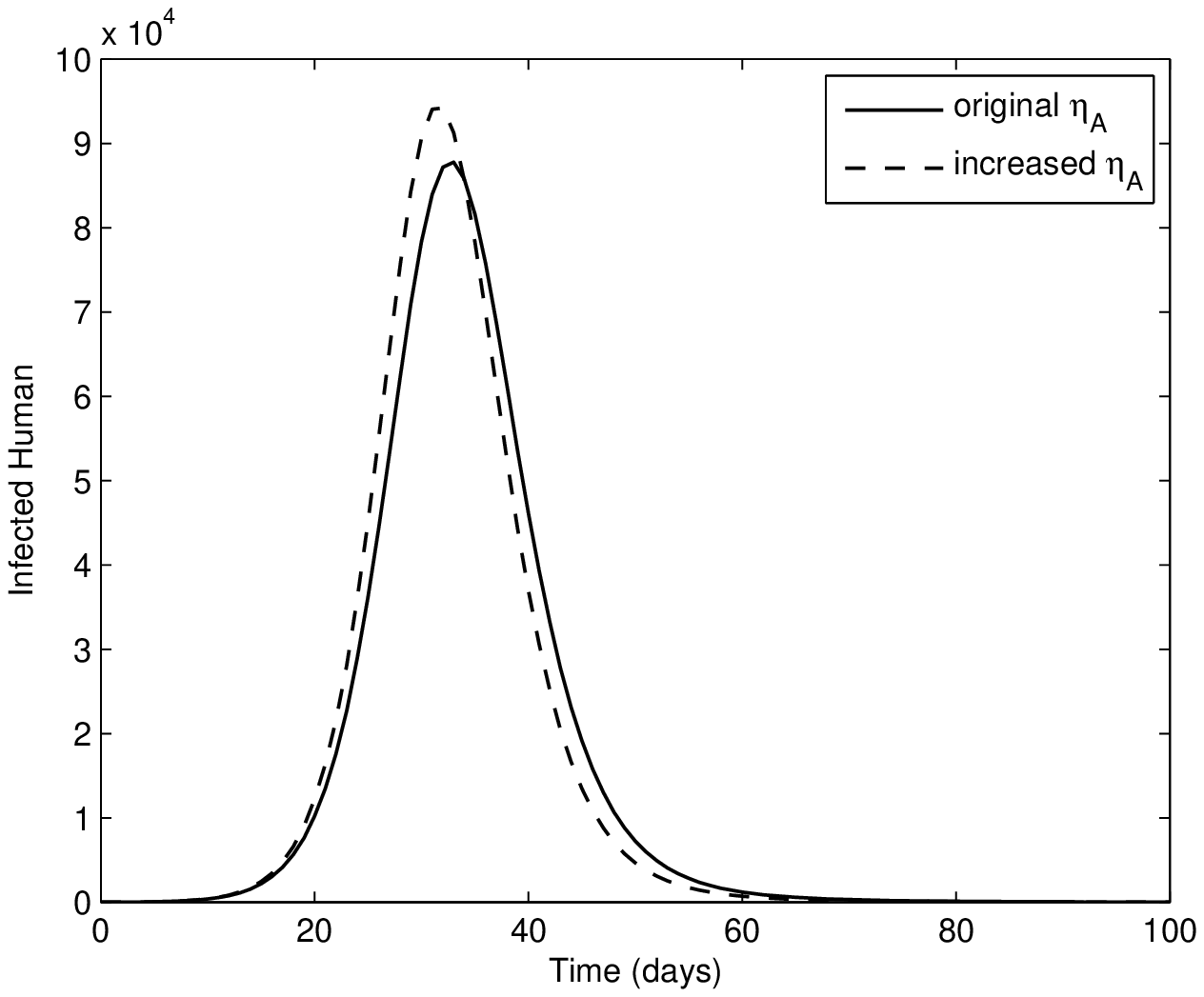}
\caption{Effect on $I_h$ of the variation of $\eta_A$.}
\label{eta_A}
\end{subfigure}
\begin{subfigure}[b]{0.45\textwidth}
\centering
\includegraphics[scale=0.44]{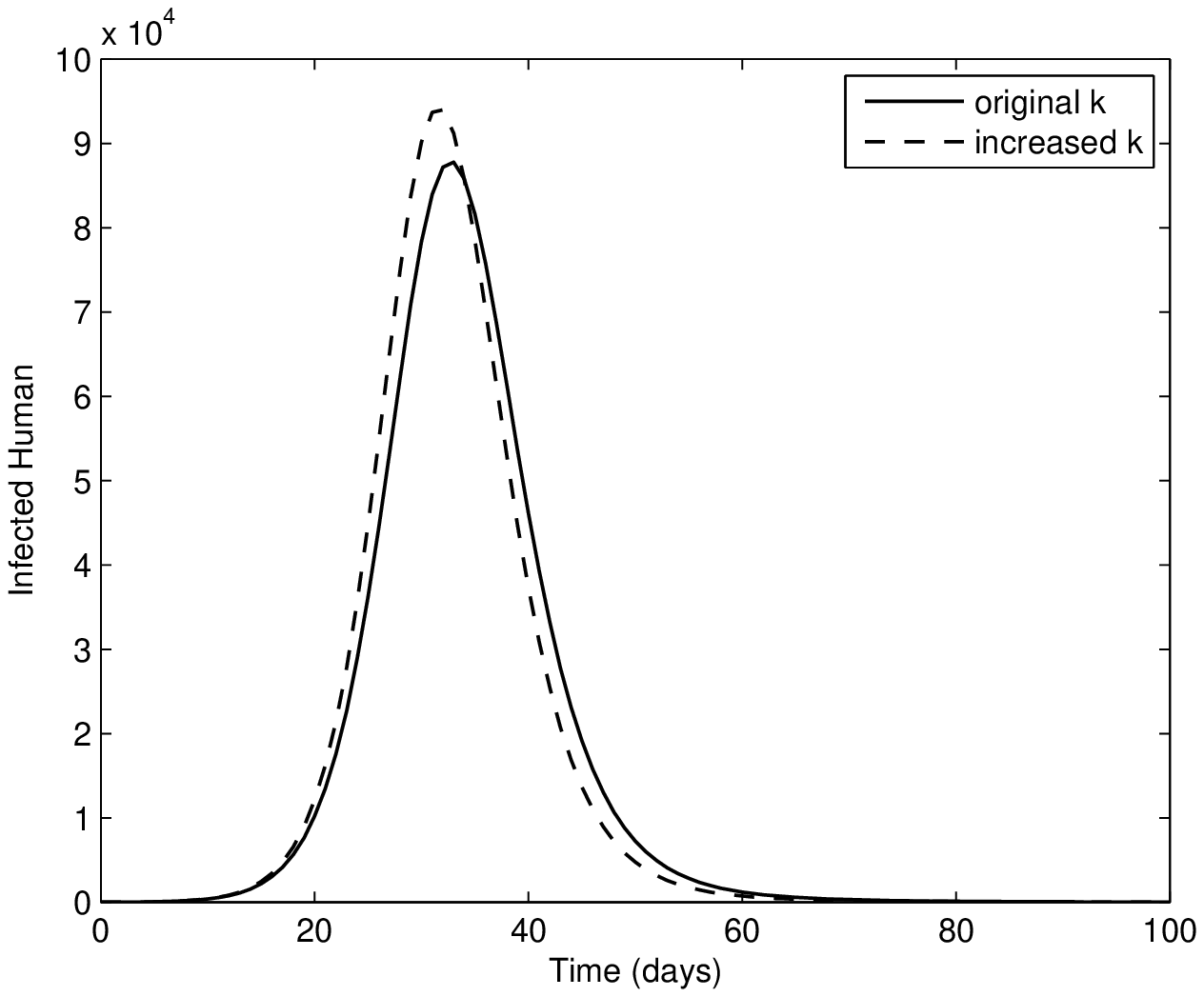}
\caption{Effect on $I_h$ of the variation of $k$.}
\label{k}
\end{subfigure}
\begin{subfigure}[b]{\textwidth}
\centering
\includegraphics[scale=0.44]{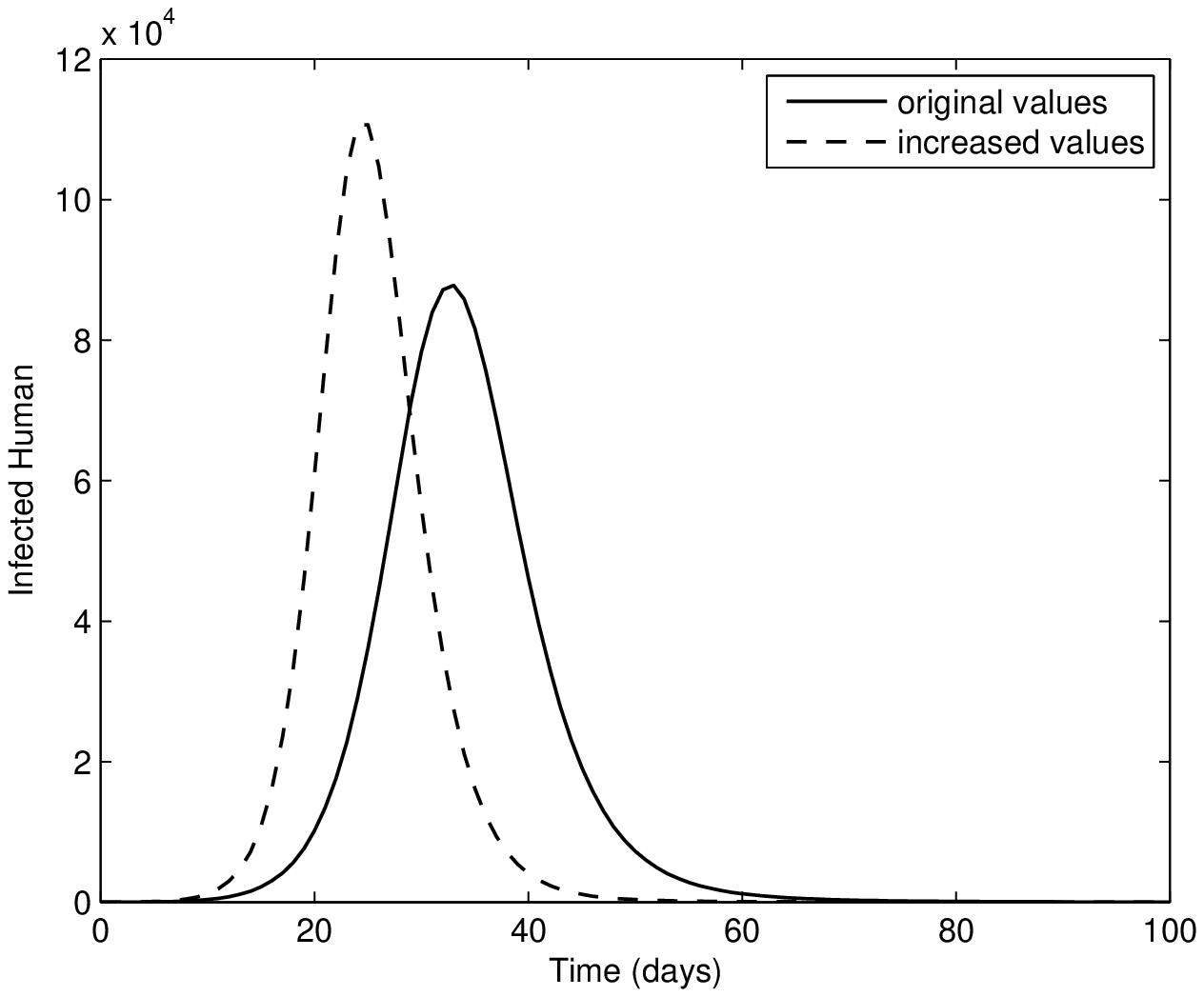}
\caption{Effect on $I_h$ when all parameters increase 10\%.}
\label{all}
\end{subfigure}
\caption{Infected individuals with initial parameter values as given in Table~\ref{parameters}
(solid line) and with an increase of 10\% of a specific (or all, in g) parameter (dashed line).}
\label{increased}
\end{figure}

Figure~\ref{increased} shows a set of graphics that reflects the effects
on the disease through parameters variation.
Each graphic presents the number of infected
humans using the baseline parameter values (solid line) described in Table~\ref{parameters}
and the corresponding curves with a specific parameter increase of $10\%$ (dashed line).

The obtained graphics reinforce the sensitivity analysis made in Section~\ref{subsection_sensitivity}.
Some parameters, $\mu_{h}$, $\varphi$ and $\mu_{A}$, present residual sensitivity indices having
small influence in  $\mathcal{R}_0$ and the changes are not graphically perceptible.
The most positive sensitive parameter is the mosquito biting rate, $B$, where
$\Upsilon_{B}^{\mathcal{R}_0}=+1$ (see Figure~\ref{B}). Figures~\ref{beta_mh}, \ref{eta_A}
and \ref{k} reflect the same behavior as the previous one with respect to
$\beta_{mh}$, $\eta_A$ and $k$ parameters, respectively. As the sensitivity index
for $\beta_{hm}$ is equal to the $\beta_{mh}$, and its effect in the infected
humans is similar, the graphic is omitted. For all these five parameters the
positive signal in the sensitivity indices of  $\mathcal{R}_0$ agrees with our intuition.

The parameters $\eta_h$ and $\mu_m$ have a negative sensitivity index.
The most negative sensitive parameter is the
average lifespan of adult mosquitoes, $\mu_{m}$, with
$\Upsilon_{\mu_{m}}^{\mathcal{R}_0}=-1.03691$.
If $\eta_h$ and $\mu_m$ are increased $10\%$, then the basic reproduction
number $\mathcal{R}_0$ decreases approximately 5\% and 10\%, respectively.
In this situation the infected humans also decrease accordingly,
as can be seen in Figures~\ref{eta_h_increased} and \ref{mu_m}.

Figure~\ref{all} presents the comparison of the infected humans when the
original parameters are considered and all the parameters are increased 10\%.

% ---------------------------------------

\section{Conclusions}
\label{sec:5}

A dengue model was studied by evaluating the sensitivity indices
of the basic reproduction number, $\mathcal{R}_0$, in order to determine
the relative importance of the model parameters in the disease transmission.
Such information allow us to identify the robustness of the model
predictions with respect to parameter values, the influence of each parameter
in the basic reproduction number, and consequently in the disease evolution.
Such analysis can provide critical information for decision makers
and public health officials, who may have to deal with the
reality of an infectious disease.

We trust that the research direction here initiated can be of great benefit to
citizens affected by dengue, with an impact on both the
prevention and control of an epidemic. Such contribution is
especially interesting regarding a disease like dengue,
which causes a large disruption in the lives of sufferers and
has enormous social and economic costs, as was well illustrated
by the outbreak of dengue that occurred in Cape Verde in 2009.

% ---------------------------------------

\section*{Acknowledgements}

This work was supported by FEDER funds through
COMPETE --- Operational Programme Factors of Competitiveness
(``Programa Operacional Factores de Competitividade'')  ---
and by Portuguese funds through the
Portuguese Foundation for Science and Technology
(``FCT --- Funda\c{c}\~{a}o para a Ci\^{e}ncia e a Tecnologia''),
within project PEst-C/MAT/UI4106/2011
with COMPETE number FCOMP-01-0124-FEDER-022690.
Monteiro was also supported by the R\&D unit Algoritmi
and project FCOMP-01-0124-FEDER-022674,
Rodrigues and Torres by the Center for Research
and Development in Mathematics and Applications (CIDMA),
Torres by project PTDC/MAT/113470/2009.

% ---------------------------------------

% ---------------------------------------

\end{document}